\documentclass[referee]{raa}            

\usepackage{graphicx,times}             
\usepackage{natbib}
\usepackage{amssymb,amsmath}
\bibpunct{(}{)}{;}{a}{}{,}
\usepackage[a4paper=true,pagebackref=true]{hyperref}
\hypersetup{colorlinks = true, linkcolor = green, anchorcolor = red, citecolor = blue, filecolor = red, pagecolor = red, urlcolor = red}

\usepackage{epstopdf}
\usepackage{makecell}
\begin{document}

   \title{A Novel Hybrid Algorithm for Lucky Imaging
\footnotetext{$\star$ Corresponding author}
}

   \volnopage{Vol.0 (20xx) No.0, 000--000}      
   \setcounter{page}{1}          

   \author{Jin-Liang Wang
      \inst{1}
   \and Bin-Hua Li
      \inst{1,2}$^\star$
   \and Xi-Liang Zhang
      \inst{3}
   }

   \institute{Faculty of information engineering and automation, Kunming University of Science and Technology, Kunming 650500, China; {\it 410784005@qq.com}, {\it lbh@bao.ac.cn}\\
        \and
             Key Laboratory of Applications of Computer Technologies of the Yunnan Province, Kunming University of Science and Technology, Kunming 650500, China\\
             \and
             Yunnan Observatories, Chinese Academy of Sciences, Kunming 650011, China\\
\vs\no
   {\small Received~~20xx month day; accepted~~20xx~~month day}}

\abstract{ Lucky imaging is a high-resolution astronomical image recovery technique with two classic implementation algorithms, i.e. image selecting, shifting and adding in image space and data selecting and image synthesizing in Fourier space. This paper proposes a novel lucky imaging algorithm where with space-domain and frequency-domain selection rates as a link, the two classic algorithms are combined successfully, making each algorithm a proper subset of the novel hybrid algorithm. Experimental results show that with the same experiment dataset and platform, the high-resolution image obtained by the proposed algorithm is superior to that obtained by the two classic algorithms. This paper also proposes a new lucky image selection and storage scheme, which can greatly save computer memory and enable lucky imaging algorithm to be implemented in a common desktop or laptop with small memory and to process astronomical images with more frames and larger size. Besides, through simulation analysis, this paper discusses the binary star detection limits of the novel lucky imaging algorithm and traditional ones under different atmospheric conditions.
}

   \authorrunning{J. L. Wang, B. H. Li \& X. L. Zhang}            
   \titlerunning{A Novel Hybrid Algorithm for Lucky Imaging }  

   \maketitle
%
%
\section{Introduction}           
\label{sect:intro}
Atmospheric turbulence, which makes the actual resolution of ground-based telescope far lower than the diffraction limited resolution, is a major factor that restricts the spatial resolution of ground-based telescope
(\citealt{Brandner+Hormuth+2016, Oscoz+etal+2008}). To obtain images with resolution close to the diffraction limited resolution of telescope in ground-based observation, either complex and expensive adaptive optics(AO) system(\citealt{Guyon+2018}) or low-cost image reconstruction technique can be used. Lucky imaging is a simple and effective astronomical image reconstruction technique to remove atmospheric turbulence effect(\citealt{Law+etal+2006, Garrel+etal+2012, Mackay+2013}). In 2003, Tubbs, Mackay and their astronomical observation group applied lucky imaging in astronomical observation and obtained very good results(\citealt{Tubbs+2003}), and from then on, many universities and scientific research institutions began the study on related algorithms. In 2006, Law et al carried out experimental study of the space-domain lucky imaging algorithm on the LuckyCam system with AO. At that time, the processing of lucky imaging (image selection, registration and addition) was carried out in image space. So the corresponding method can be called space-domain lucky imaging (LI). In 2012, Garrel, Guyon and Baudoz(\citealt{Garrel+etal+2012})proposed the a new lucky imaging algorithm that selects good data and synthesizes an image in Fourier space and carried out corresponding experimental study with AO system-based simulated data. It is a lucky imaging method in frequency domain and called lucky Fourier (LF) or GGB method by \cite{Mackay+2013}. In 2013, Mackay (\citealt{Mackay+2013})proposed a hybrid algorithm with the low spatial frequency parts selected according to the LF procedure and the high spatial frequency parts selected essentially from the classic LI procedure and obtained better results. But this method suffers two drawbacks: the size of the central low spatial frequency patch is selected by trial and error, and the imaging effect is dependent on the patch and the given selection percentages. In 2018, Mao et al subjected the conventional lucky imaging algorithm to partial graphics processing unit (GPU) acceleration (\citealt{Mao+etal+2018})while Zhao et al (\citealt{Zhao+etal+2019})used field programmable gate array (FPGA) for experimental study of the lucky imaging algorithm; in 2019, Hu et al (\citealt{Hu+2019})carried out experimental study of packet processing of the LF algorithm on an observation system without AO; and in 2020, Wang et al (\citealt{Wang+etal+2020})modified the conventional lucky imaging algorithm, proposed a algorithm where real-time processing is possible, and carried out on a FPGA system the corresponding experimental study.
\par Under normal circumstances, a certain selection percentage is used for image selection and processing in both space-domain and frequency-domain lucky imaging, generally 1\% for the former(\citealt{Oscoz+etal+2008, Law+etal+2006, Mao+etal+2018})and 10\% for the latter(\citealt{Garrel+etal+2012, Hu+2019}). Different selection rates can bring different results of lucky imaging, so the algorithm with a given selection rate may miss a lot of important information and is difficult to guarantee that the final high-resolution reconstructed image is the optimal. Meanwhile, storage of all astronomical images in space-domain and frequency-domain lucky imaging leads to large occupation of computer memory, so these two algorithms are neither suitable to be implemented on common desktop or laptop with small memory, nor to be used to process astronomical images with more frames and larger size. 
\par On the basis of the study by Mao and Hu from our laboratory, this paper proposes a novel hybrid lucky imaging algorithm, where with selection rate, the key factor for both LI and LF algorithms, as a link, the two classic algorithms are combined as its proper subsets. Meanwhile, the selection rates used in LI and LF algorithms are made an increasing sequence instead of fixed values, good images or data are selected and superposed in many levels and aspects according to certain rules (e.g. the instantaneous Strehl ratio for LI algorithm and the amplitude for LF algorithm), and a high-resolution reconstructed image is obtained through fusion of various results. Finally, a large amount of short-exposure images taken by the 2.4m-diameter astronomical telescope at the Lijiang Observatory, Yunnan Observatories, Chinese Academy of Sciences are used to verify the feasibility of the hybrid algorithm and the experiment results of the algorithm are analyzed and discussed. This paper also proposes a new image selection and storage scheme for lucky imaging algorithm, which can greatly save computer memory and enable lucky imaging algorithm to be implemented in a common desktop or laptop with small memory and to process astronomical images with more frames and larger size. 
\par The main contributions of this study include the following three aspects:
\par (1)	The authors propose a novel hybrid algorithm of lucky imaging in image space and Fourier space.
\par (2)	The authors propose a new image selection and storage scheme.
\par (3)	Through simulation analysis, the authors discuss the binary star detection limits of the proposed algorithm and traditional ones under different atmospheric conditions.
\par In Section 2, we introduce the principle of the novel hybrid lucky imaging algorithm, the new image selection and storage scheme, the algorithm flow. Verification and Comparative performances of the proposed algorithm based on observed and simulated imaged are discussed in Section 3. The conclusions are given in Section 4.
\section{Hybrid Lucky Imaging Algorithm}
\label{sect:Obs}

The basic principle of lucky imaging algorithm is to get high-resolution images with resolution close to the diffraction limited resolution of telescope through a series of processing of effective information obtained from a large number of short-exposure images. There are two basic algorithm implementation frameworks: one is space-domain lucky imaging algorithm (also called classic lucky imaging algorithm), of which the basic procedures are to carry out image quality evaluation of short-exposure images (the instantaneous Strehl ratio, i.e. the peak of each frame, is generally used as the image quality evaluation function of a point target) and then to select at a certain percentage (i.e. selection rate) a number of high-quality images for registration and superposition, thus reconstructing high-resolution images(\citealt{Law+etal+2006, Mao+etal+2018}); and the other is frequency-domain lucky imaging algorithm, of which the basic procedures are to register in space domain the short-exposure images with the peak of each frame as the center to recenter and clip the images, then to subject the clipped images to fast Fourier transform (FFT) and save their complex values ( real parts and imaginary parts) and the corresponding amplitudes, to select at a certain percentage a number of complex values from different frames with the maximum amplitude at each spatial frequency in Fourier plan and sum the chosen complex values, and finally to get high-resolution images through the inverse FFT (IFFT)(\citealt{Garrel+etal+2012, Hu+2019}). Both the two algorithms can make distinguishable the bright and faint stars that cannot be otherwise distinguished and resolvable the details that cannot be otherwise resolved. However, selection rate, a key factor for both the two algorithms, has a significant influence on the result of lucky imaging. In this paper, a novel lucky imaging algorithm is obtained through fusion of the two classic algorithms with the selection rate as a link. An introduction is made below to the principle of the novel hybrid algorithm, then to a new scheme for image/data selection and storage, and finally to the procedures of the proposed algorithm. 

\subsection{Principle of Hybrid Lucky Imaging Algorithm}

In the conventional LI algorithm, good images are selected in space domain ( image space) at a certain selection rate for registration and superposition, and in the LF algorithm, good data (complex values) are selected in frequency domain (Fourier space) at a certain selection rate at each spatial frequency for superposition. Both algorithms involve data selection and superposition in their respective processing domains under the control of their certain selection rates. Space-domain processing algorithm is simpler, with lower memory consumption but lower information utilization; frequency-domain processing algorithm has better effect, with higher information utilization but higher calculation amount and memory consumption. Besides, selection rate has a significant influence on the results of both algorithms. According to the study by Garrel et al and that by Mackay, the imaging result of frequency-domain algorithm at a 10\% selection is almost the same as that of space-domain algorithm at a 1\% selection(\citealt{Garrel+etal+2012, Mackay+2013}). Regardless of whether it is a lucky imaging in space domain or frequency domain, the basic idea is to extract high-resolution information from a large number of short-exposure images at a selection rate in its own processing domain, and to reconstruct an astronomical image with this information. Therefore, the reconstruction result is directly related with the value of the selection rate; that is, the selection rate will affect the lucky imaging result.

The motivation of this paper is to develop a novel algorithm for lucky imaging that does not depend on a specific selection rate; that is, the algorithm can extract the effective high-resolution information from short-exposure images to the maximum, and generate a resultant image independent of the specific selection rate. The lucky imaging results obtained with different selection rates contain different effective information. For this reason, we do not perform lucky imaging in spatial or frequency domain alone, but combine the space-domain and frequency-domain selection rates from the original two independent one-dimensional point sets on axes into a two-dimensional point set on a plane. For each point on the plane, the LI and the LF are combined to generate one image. For all the images obtained in this way, the effective information is extracted again according to the frequency-domain lucky imaging rule; that is, the second frequency-domain fusion is performed to obtain the final lucky imaging result. In this way, the information utilization of the original short exposure images is improved, and the final result is independent of the specific selection rate.

\begin{figure}[ht]
   \centering
   \includegraphics[width=5in, angle=0]{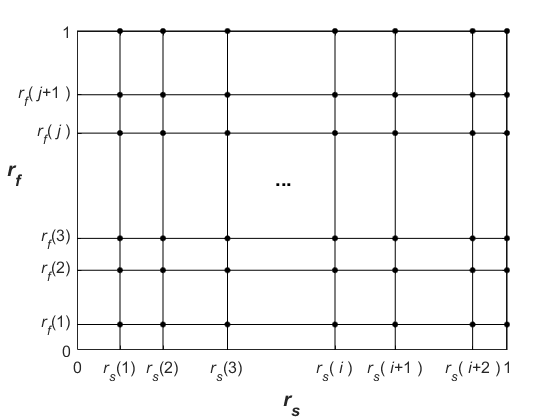}
   \caption{Schematic diagram of the hybrid lucky imaging algorithm. }
   \label{Fig1}
   \end{figure}

The basic solution is to select different image sets in image space for image fusion at different complex data selection rates in Fourier space. To obtain different image sets and better imaging effect, the percentage for selection of image sets in image space is variable, which is the same for image selection in classic LI algorithm. As the complex data selection rate is also variable in LF algorithm, multiple fused images can be formed in Fourier space. Finally, complex values at each spatial frequency are selected again in these fused images according to amplitude rule to form a final fused Fourier image. For the schematic diagram of the hybrid algorithm, see Figure~\ref{Fig1}. 

In Figure~\ref{Fig1}, abscissa $r_s$ is the space-domain selection rate and ordinate $r_f$ is the frequency-domain selection rate; the set of black points on line $r_f = 1$ and that on line $r_s = 1$ are the point set of space-domain LI algorithm and that of frequency-domain LF algorithm respectively, and the set of crossed black points on plane $(r_s, r_f)$ is the point set of the hybrid algorithm of LI and LF. The grid spacing can be uneven, for the increment of each step of space-domain or frequency-domain selection rate can be different.

The initial point and terminal point of the hybrid algorithm are $(r_s(1), r_f(1))$ and $(1,1)$ respectively. The iterative formula of the space-domain selection rate is as shown in Equation~\ref{eq:LebsequeI1}:
\begin{equation}
  r_s(i) = r_s(i-1) + d_s(i-1), ~~~~~~i > 1
\label{eq:LebsequeI1}
\end{equation}
where $d_s$ is the increment of the space-domain selection rate. The iterative formula of the frequency-domain selection rate is as shown in Equation~\ref{eq:LebsequeI2}: 
\begin{equation}
  r_f(j) = r_f(j-1) + d_f(j-1), ~~~~~~j > 1
\label{eq:LebsequeI2}
\end{equation}
where $d_f$ is the increment of the frequency-domain selection rate. The iterative formula of the selection rate $r_{sf}$ for the hybrid algorithm is as shown in Equation~\ref{eq:LebsequeI3}: 
\begin{equation}
  r_{sf}(i,j) = r_s(i) \times r_f(j), ~~~~~~i,j > 1
\label{eq:LebsequeI3}
\end{equation}
The number of the selected points of the space-domain selection rate $p_s$ is the length of sequence $r_s$, as shown in Equation~\ref{eq:LebsequeI4}:
\begin{equation}
  p_s = length(r_s)
\label{eq:LebsequeI4}
\end{equation}
The number of the selected points of the frequency-domain selection rate $p_f$ is the length of sequence $r_f$, as shown in Equation~\ref{eq:LebsequeI5}:
\begin{equation}
  p_f = length(r_f)
\label{eq:LebsequeI5}
\end{equation}
The number of the selected points of the hybrid algorithm $p_{a}$ is the product of the above two, i.e. 
\begin{equation}
  p_{a} = p_s \times p_f
\label{eq:LebsequeI6}
\end{equation}
Specific implementation method: First, set four parameters $r_s(1)$ and $d_s$, $r_f(1)$ and $d_f$, respectively the initial value and the increment of the space-domain selection rate and those of the frequency-domain selection rate; then, form in an iterative way an increasing sequence of the space-domain selection rates and an increasing sequence of the frequency-domain selection rates, with terminal values of both sequences being 1, i.e. both the final space-domain and the final frequency-domain selection rates being 1; next, for each selection rate in image space, select goods images according to the instantaneous Strehl ratio and convert them into the corresponding Fourier images by FFT; afterward, for each selection rate in Fourier space, select and superpose good complex data according to the index of maximum amplitude (i.e. amplitude rule); finally, for $p_{a}$ Fourier images, select the complex data with the largest amplitude at each spatial frequency to synthesize a final Fourier image, and use IFFT to get the final high-resolution reconstructed image.

\subsection{Scheme for Image/Data Selection and Storage}

No matter in space-domain or frequency-domain lucky imaging algorithm, image or complex data selection and storage is the most critical step. In traditional LI algorithm, the scheme for image selection is to carry out comparison and sorting based on the instantaneous Strehl ratio of each frame and that for image storage is to store all images in computer memory(\citealt{Law+etal+2006, Mao+etal+2018}) while in LF algorithm, the scheme for complex data selection is to carry out comparison and sorting based on the amplitude of each Fouier image at each spatial frequency and that for complex data storage is to store all Fourier images and their amplitudes in computer memory(\citealt{Garrel+etal+2012, Mackay+2013, Hu+2019}). When there are a large number of images, the selection algorithm based on comparison and sorting will require a large amount of calculation and occupy a huge amount of computer memory, so they are neither suitable to be implemented on common desktop or laptop with small memory, nor to be used to process astronomical images with more frames and larger size. This problem is more remarkable in LF algorithm. Therefore, considering that the sequence of selection results has no impact on subsequent operations, this paper adopts the idea of real-time image selection in image space based on comparison but no sorting as proposed in Ref.\cite{Wang+etal+2020} to design applicable selection and storage schemes for LI and LF algorithms which can greatly save computer memory. 

\begin{figure}[ht]
   \centering
   \includegraphics[width=5in, angle=0]{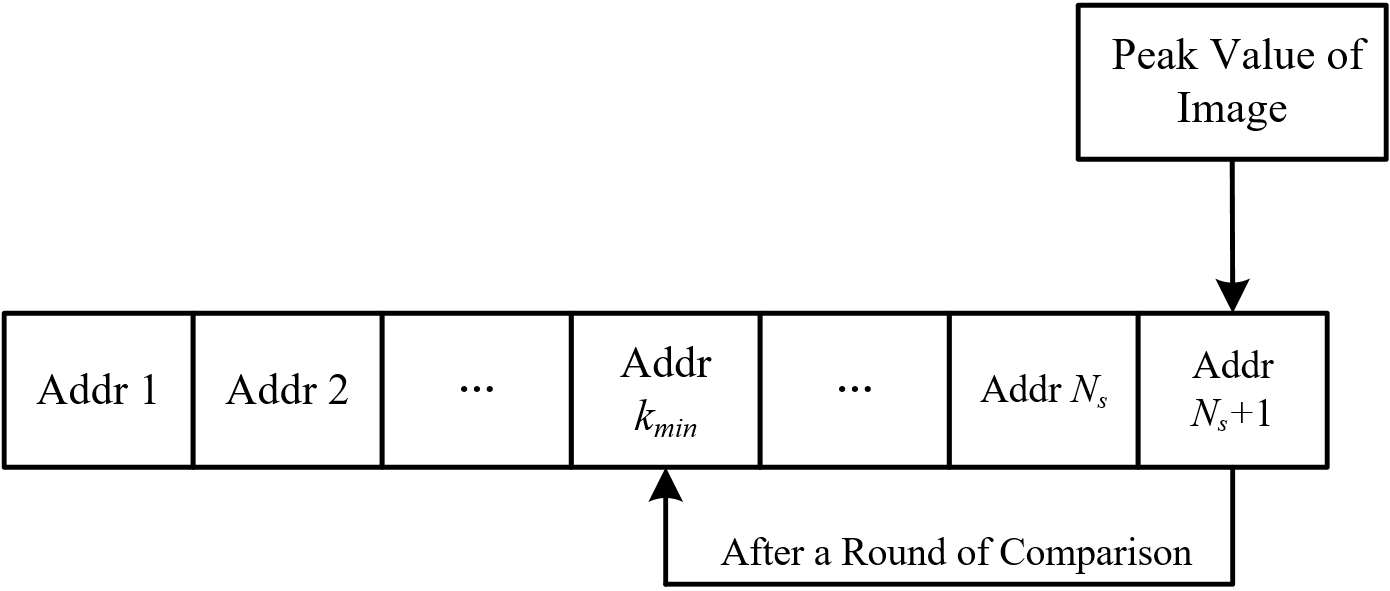}
   \caption{Schematic diagram of the image selection and storage scheme. }
   \label{Fig2}
   \end{figure}

In this new image selection and storage scheme for LI algorithm, the peak of the current input image replaces the minimum of the selected peaks which is found through comparison and the current clipped(i.e. shifted and resized) image replaces the clipped image corresponding to the minimum in the image storage area. Its implementation diagram is as shown in Figure~\ref{Fig2}. 

In Figure~\ref{Fig2}, $N_s$ is the number of selected images (the product of total frame number $N$ and space-domain selection rate $r_s$), $k_{min}$ is the address unit where the minimum is located. It can be seen from Figure~\ref{Fig2} that this image selection and storage scheme can guarantee that the selected peak data are the first $N_s$ maximums of $N$ peaks of frames and that the data in the storage region of clipped images are the data of clipped images with the top $N_s$ maximums, thus achieving the goal of selecting a number of images with the maximum peaks at a certain space-domain selection rate. 

In the new complex data selection and storage scheme for LF algorithm, the current Fourier amplitude corresponding to each spatial frequency replaces the minimum amplitude which is found through comparison and the current complex data are written to the complex data unit corresponding to the minimum. Its implementation diagram is largely consistent with Figure~\ref{Fig2} except that the input is the amplitude corresponding to each spatial frequency and that $m \times n$ cycles are required as each frame of the clipped image has $m \times n$ spatial frequencies ($m$ and $n$ are respectively the line number and the column number of clipped image). Similarly, this scheme can guarantee that the data in address unit 1 to address unit $N_f$ of each line are the top $N_f$ maximums of $N$ amplitudes of each line and it is the same for the corresponding complex data, where $N$ is the total frame number and $N_f$ is the number of selected Fourier images (the product of total frame number $N$ and frequency-domain selection rate $r_f$), thus achieving the goal of selecting a number of complex data with the maximum amplitudes at a certain frequency-domain selection rate. 

\subsection{Hybrid Lucky Imaging Algorithm Flow}

\begin{figure}[ht]
   \centering
   \includegraphics[width=5in, angle=0]{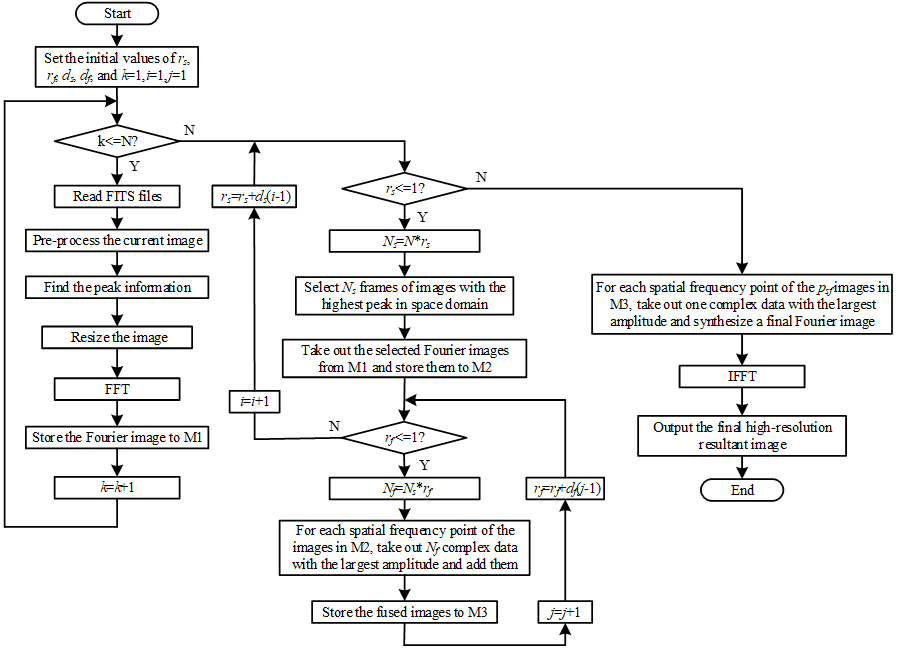}
   \caption{Algorithm flowchart. }
   \label{Fig3}
   \end{figure}

With the above-mentioned image selection and storage scheme for LI algorithm and amplitude selection and storage scheme for LF algorithm embedded into the process of space-domain and frequency-domain lucky imaging, a novel hybrid lucky imaging algorithm is formed, its flow being described in Figure~\ref{Fig3}, in which M1, M2 and M3 represent three different memory areas.

It can be seen from the above algorithm flow that the hybrid algorithm, with space-domain and frequency-domain selection rates as a link, skillfully combines the LI and LF algorithms as its proper subsets. When frequency-domain selection rate is 1, the hybrid algorithm degrades into a space-domain LI algorithm; when space-domain selection rate is 1, the hybrid algorithm degrades into a frequency-domain LF algorithm. 

\section{Experimental Results and Analysis}
\label{sect:data}

\subsection{Trial Datasets and Computing Platform}
This experiment is based on the data observed with the 2.4m general-purpose optical telescope at Lijiang Observatory of Yunnan Observatories. Since it is an experimental observation, two brighter binary systems HDS 70 and TOK 382 are selected as the targets. The relevant parameters of the two targets are downloaded from Double Star Database of Stelle Doppie\inst{1}, shown in Table~\ref{Tab1:publ-works}.

\footnotetext[1]{https://www.stelledoppie.it/index2.php?}

\begin{table}
\begin{center}
\caption[]{  HDS 70 and TOK 382.}\label{Tab1:publ-works}


 \begin{tabular}{clclclcl}
  \hline\noalign{\smallskip}
\makecell[c]{WDS\_NAME} &  \makecell[c]{LAST}      & \makecell[c]{PA} & \makecell[c]{SEP} & \makecell[c]{MAG1}  & \makecell[c]{MAG2} & \makecell[c]{D\_MAG}                   \\
  \hline\noalign{\smallskip}
\makecell[c]{HDS 70}  & \makecell[c]{2010} & \makecell[c]{66}     & \makecell[c]{0.30} & \makecell[c]{7.84} & \makecell[c]{9.37} & \makecell[c]{1.53}  \\ 
\makecell[c]{TOK 382}  & \makecell[c]{2018} & \makecell[c]{204}     & \makecell[c]{0.40} & \makecell[c]{4.70} & \makecell[c]{8.70} & \makecell[c]{4.00}                  \\
  \noalign{\smallskip}\hline
\end{tabular}
\end{center}
\vspace{-0.8cm}
\end{table}

Optical system parameters: telescope aperture 2.4m, filter center wavelength 850nm, bandwidth 50nm, magnification 4, focal length 29.1m. An EMCCD camera (Andor iXon 897) is used as the imaging terminal. The EM gain is set to 300, the exposure time is 20ms. The observation date is October 20, 2016. The weather is clear at night, with an average wind speed of 5.27km/h and an average seeing of 1.4". Three sets of images were taken for each target, each of 10,000 frames. The image size of each frame is $512 \times 512$ pixels, and the CCD image scale is 0.043"/pixel.

\begin{figure}[ht]
\centering
  \subfloat[A random frame of HDS 70.]{
  \label{fig1:subfig:a}
  \begin{minipage}[t]{0.5\linewidth}
  \centering
   \includegraphics[scale=0.9]{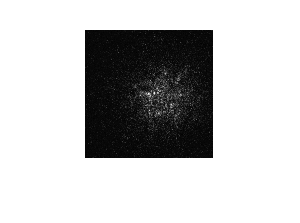}
  \end{minipage}%
  \begin{minipage}[t]{0.5\linewidth}
  \centering
   \includegraphics[scale=0.45]{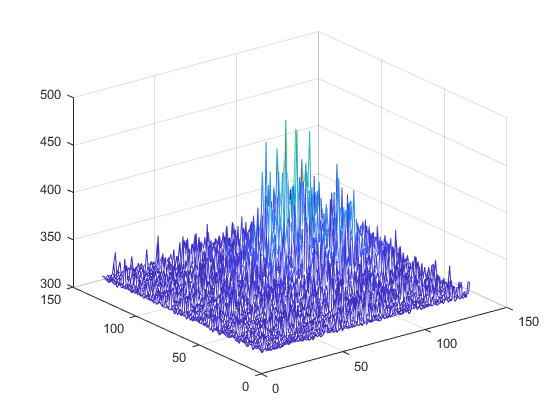}
  \end{minipage}%
  }
  
  \subfloat[A random of TOK382.]{
  \label{fig1:subfig:b}
  \begin{minipage}[t]{0.5\linewidth}
  \centering
   \includegraphics[scale=0.9]{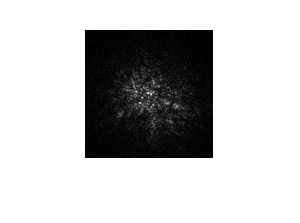}
  \end{minipage}%
  \begin{minipage}[t]{0.5\linewidth}
  \centering
   \includegraphics[scale=0.45]{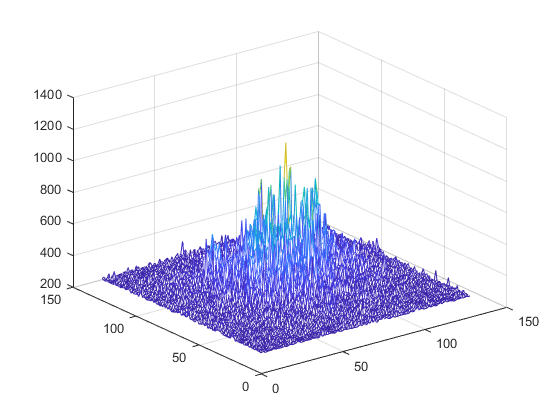}
  \end{minipage}%
  }
  \centering
  \caption{Observed images of binary HDS 70 and TOK382.}
  \label{Fig:fig1}
\end{figure}

In this paper, one of the three sets is randomly selected for algorithm verification and the other two for auxiliary verification. Observed images and their 3D gray value distribution of HDS70 and TOK382 are shown in Figure~\ref{fig1:subfig:a} and \ref{fig1:subfig:b} respectively.

The computing platform of the hybrid lucky imaging algorithm is a Dell Precision T5500 imaging workstation, of which the memory is 16GB, the CPU is Xeon E5620, the graphics card is NVIDIA GTX1080, the software environment is Windows 7 operating system and MATLAB R2017a, and the algorithm is programmed and implemented by MATLAB.

\subsection{Verification of Hybrid Lucky Imaging Algorithm}

\begin{figure}[ht]
\centering
  \subfloat{
  \begin{minipage}[t]{0.5\linewidth}
  \centering
   \includegraphics[scale=0.9]{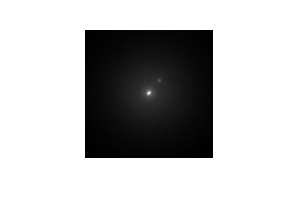}
  \end{minipage}%
  \begin{minipage}[t]{0.5\linewidth}
  \centering
   \includegraphics[scale=0.45]{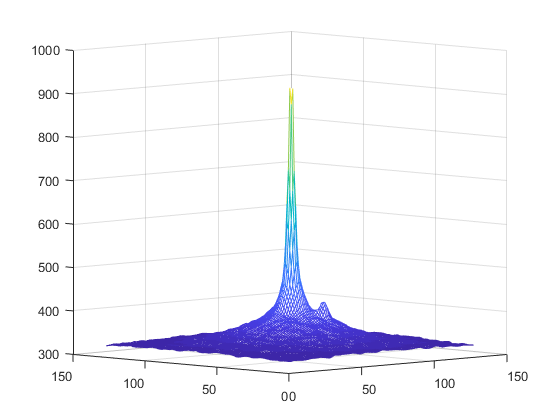}
  \end{minipage}%
  }
   \centering
  \caption{Final resultant images.}
  \label{Fig5}
\end{figure}

This experiment is to process the observed image as per the algorithm flow described in Section 2.2 and obtain a high-resolution reconstructed image. The image used in the experiment is of $512 \times 512$ pixels in size, of totally 10,000 frames, and is resized or clipped into $128 \times 128$ pixels with the peak as the center. Such clipping is adopted as it can reduce the size of reconstructed area in which the observation objects located, the occupation of computer memory and the volume of data to be processed, thus improving processing speed. Therefore, in this experiment, $m=n=128$, $N=10,000$, and the initial value and the increment in algorithm flow can be set at will to a value between 0 and 1; the more terms of increasing sequence there are in horizontal and vertical directions, the more detailed information of image can be captured and the better effect can be obtained while the longer running time will be required. Giving the compromise between effect and running time, the initial value $r_s(1)$ and the increment $d_s$ of space-domain selection rate and the initial value $r_f(1)$ and the increment $d_f$ of frequency-domain selection rate are all set to 0.1 in the algorithm flow, in which case $p_s= p_f =10$, and therefore, after processing by lucky imaging algorithm at different space and frequency-domain selection rates, totally $p_{a}=100$ resultant images are obtained, which are subjected to fusion in frequency domain upon the maximum amplitude rule at each spatial frequency and then to IFFT to get the final high-resolution reconstructed image. As a large diffuse halo around the bright primary of the binary are likely to cover the rays of the faint star and even change its gray value, therefore, in Figure~\ref{Fig5}, a logarithmic transformation is performed on the two-dimensional gray image, so as to make the faint companion easy to be distinguished.

The targets (bright primary and faint companion) are subjected to centering calculation by two-dimensional-modified moment centering algorithm(\citealt{Ji+Wang+1996}) to obtain the angular separation between the primary and the companion, approximately 0.328 arcsecond, which is close to the result given in the paper by \cite{Mao+etal+2018}. Besides, through simple diameter photometry, it can be concluded that the ratio of brightness (sum of pixel values minus background) of primary to that of companion is approximately 4.37 (corresponding magnitude difference of approximately 1.60mag), which deviates slightly from the actual brightness ratio of primary and companion of HDS70, approximately 4.09 (corresponding magnitude difference of 1.53mag, shown in Table~\ref{Tab1:publ-works}). The possible reason for this deviation may be that our measurement method is too simple. In fact, the luminous flux measurement of Lucky Imaging is highly related to the selection percentage (\citealt{Mackay+2013}), so the photometric accuracy is not too high. Therefore, viewed from the visual effect of resultant image, the separation of binary stars and the magnitude difference, the lucky imaging results obtained by this algorithm are reliable. 

\begin{figure}[ht]
   \centering
   \includegraphics[width=5in, angle=0]{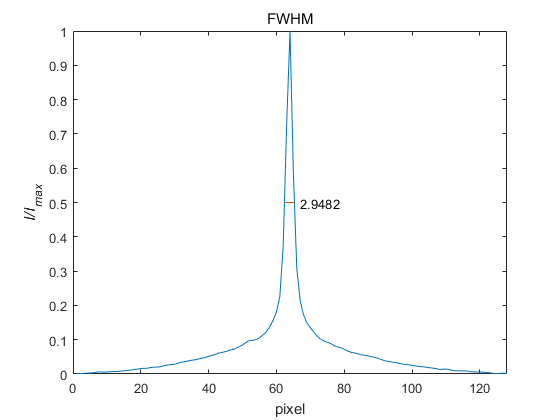}
   \caption{Profile and FWHM of bright star. }
   \label{Fig6}
   \end{figure}

The resolution of astronomical images of stars is mainly measured by angular resolution and generally represented by the full width at half maximum (FWHM) of bright star. Therefore, to test the resolving power of the experimental system, an average value of four sections of high-resolution images of binary HDS70 (i.e. the horizontal direction, the vertical direction, and the two diagonal directions) is calculated, and then the average section is fitted with a piecewise curve to obtain the contour curve of the bright star, as shown in Figure~\ref{Fig6}, from which it is concluded that the FWHM is 2.9482 pixels; then, based on the scale of the telescope imaging system of 0.043"/pixel, it is concluded through conversion that the FWHM is approximately 0.127", which is close to the limiting resolution of 2.4m-diameter telescope (0.09") and far lower than the FWHM of image subjected to long exposure at the night of observation (approximately 1.4"), indicating that the experimental system can indeed improve significantly the resolution of the ground-based telescope.

Furthermore, we have tested the other two datasets of observed images (each of 10,000 frames) of HDS 70 and obtained similar results, indicating that the hybrid lucky imaging algorithm proposed in this paper is correct and feasible. 

\subsection{Comparative Analysis}   
Owning to the continuity of our study work, the experimental data and platform for the hybrid lucky imaging algorithm proposed in this paper are completely the same as those for the space-domain LI algorithm designed and implemented by Mao and those for the frequency-domain LF algorithm by Hu while totally different from those for the lucky imaging algorithm studied by Mackay and that by Garrel et al, so the algorithm implemented by Mao and that by Hu are the main objects of comparative analysis and discussion in our study process.

First, we compare the selection and storage scheme introduced in Section 2.2 with those designed and implemented by Mao and by Hu. Taking 10,000 frames of HDS70 images with $512 \times 512$ pixels as the processing object, the memory consumption of the four different algorithms in MATLAB computing environment is shown in Table~\ref{Tab2:publ-works}. The traditional LI and traditional LF are the traditional space-domain and frequency-domain lucky imaging algorithms designed and implemented by Mao and by Hu using MATLAB, respectively; and the Improved LI and Improved LF are the modified algorithms of traditional LI and LF algorithm according to the selection and storage scheme proposed in this paper, respectively. It can be seen from Table~\ref{Tab2:publ-works} that the total memory required by the new scheme is only 9.59\% of that of the traditional scheme for the space-domain algorithm at the selection rate of 1\%, and 31.35\% of the traditional scheme for the frequency-domain algorithm at the selection rate of 10\%. In fact, Mao's and Hu's schemes are consistent with those for the classic LI and LF algorithms, in which sorting-based selection and all storage are adopted, occupying a large amount of computer memory. It can be seen from Section 2.2 that in the new scheme, only the data selected according to certain rules (e.g. the instantaneous Strehl ratio for LI algorithm and the amplitude for LF algorithm) rather than all data will be stored, so this scheme can significantly save computer memory and enable lucky imaging algorithm to be implemented in a common desktop or laptop with small memory and to process astronomical images with more frames and larger size.

\begin{table}
\begin{center}
\caption[]{ Memory required for LI and LF algorithms based on 2 different selection and storage schemes.}\label{Tab2:publ-works}


 \begin{tabular}{clcl}
  \hline\noalign{\smallskip}
\makecell[c]{Algorithm} &  \makecell[c]{Physical Memory/MB}                    \\
  \hline\noalign{\smallskip}
\makecell[c]{Traditional LI}           & \makecell[c]{1711}                  \\
\makecell[c]{Improved LI}          & \makecell[c]{164}                  \\
\makecell[c]{Traditional LF}           & \makecell[c]{7481}
\\
\makecell[c]{Improved LF}            & \makecell[c]{2345}                  
        \\
  \noalign{\smallskip}\hline
\end{tabular}
\end{center}
\vspace{-0.8cm}
\end{table}

Then, for resultant images of different lucky imaging algorithms, we select 7 objective image quality (definition) assessment functions, i.e. FWHM, Brenner gradient, Tenengrad gradient, Laplacian gradient, sum of modulus of gray difference (SMD), gray variance, energy of gradient (EOG), and make comparative analysis and discussion through numerical calculation. In these 7 assessment functions, FWHM is a commonly used assessment index for astronomical images and the lower the FWHM is, the better the image quality is; other 6 functions are all definition assessment indexes for classic digital images and the higher their values are, the better the image quality is. The quality assessment results of three reconstructed resultant images obtained respectively through the LI algorithm designed and implemented by Mao, the LF algorithm by Hu, and the proposed hybrid algorithm are shown in Table~\ref{Tab3:publ-works}.

\begin{table}
\begin{center}
\caption[]{ Quality Assessment Results of Images Obtained Through Three Algorithms.}\label{Tab3:publ-works}


 \begin{tabular}{clcl}
  \hline\noalign{\smallskip}
\makecell[c]{Assessment}       & \makecell[c]{LI(\citealt{Mao+etal+2018})} & \makecell[c]{LF(\citealt{Hu+2019})}      &  \makecell[c]{\textbf{Proposed}}              \\
  \hline\noalign{\smallskip}
\makecell[c]{FWHM}   & \makecell[c]{0.202"}     & \makecell[c]{0.171"} & \makecell[c]{\textbf{0.127"}} \\ 
\makecell[c]{Brenner}       &   \makecell[c]{11.99}     & \makecell[c]{33.84}   & \makecell[c]{\textbf{54.43}}               \\
\makecell[c]{Tenengrad}       &   \makecell[c]{4.35}     & \makecell[c]{9.51}          & \makecell[c]{\textbf{17.12}}        \\
\makecell[c]{Laplacian}       &   \makecell[c]{3.47}     & \makecell[c]{9.73} & \makecell[c]{\textbf{10.60}}
\\
\makecell[c]{SMD}      &   \makecell[c]{20.67}     & \makecell[c]{24.98}          & \makecell[c]{\textbf{37.74}}         \\
\makecell[c]{Variance}       &   \makecell[c]{157.37}     & \makecell[c]{158.44}         & \makecell[c]{\textbf{233.82}}         \\
\makecell[c]{EOG}       &   \makecell[c]{9.82}     & \makecell[c]{39.73}  & \makecell[c]{\textbf{47.88}}
        \\
  \noalign{\smallskip}\hline
\end{tabular}
\end{center}
\vspace{-0.8cm}
\end{table}

It can be seen from Table~\ref{Tab3:publ-works} that the FWHM of the proposed hybrid lucky imaging algorithm is lower than that of the LI algorithm designed and implemented by Mao and that of the LF algorithm by Hu, indicating that the quality of the resultant image obtained through the hybrid algorithm is better than that obtained through LI and LF algorithms. As to the other 6 indexes, the values for LF algorithm are all larger than those for LI algorithm, indicating that frequency-domain algorithm is superior to space-domain algorithm, and the values for the proposed algorithm are all larger than those for frequency-domain algorithm, indicating that the results of this algorithm are the best; that is to say, the quality of the resultant image obtained through the hybrid algorithm is the highest. 

\begin{figure}[ht]
\centering
  \subfloat[Simple addition of all images.]{
  \label{fig2:subfig:a}
  \begin{minipage}[t]{0.5\linewidth}
  \centering
   \includegraphics[scale=0.8]{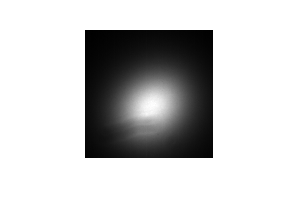}
  \end{minipage}%
  \begin{minipage}[t]{0.5\linewidth}
  \centering
   \includegraphics[scale=0.4]{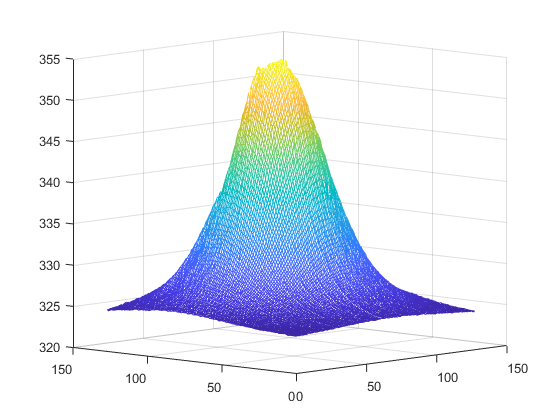}
  \end{minipage}%
  }
  
  \subfloat[Shift and addition of all images.]{
  \label{fig2:subfig:b}.
  \begin{minipage}[t]{0.5\linewidth}
  \centering
   \includegraphics[scale=0.8]{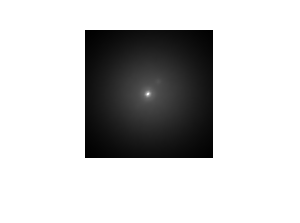}
  \end{minipage}%
  \begin{minipage}[t]{0.5\linewidth}
  \centering
   \includegraphics[scale=0.4]{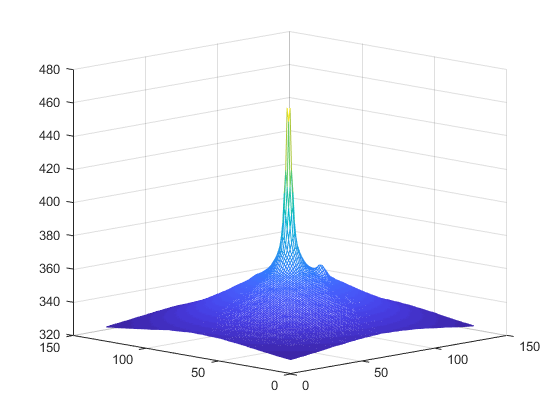}
  \end{minipage}%
  }
  
    \subfloat[LI.]{
  \label{fig2:subfig:c}
  \begin{minipage}[t]{0.5\linewidth}
  \centering
   \includegraphics[scale=0.8]{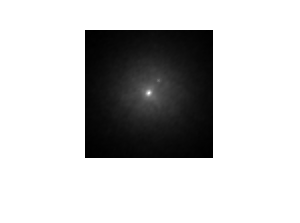}
  \end{minipage}%
  \begin{minipage}[t]{0.5\linewidth}
  \centering
   \includegraphics[scale=0.4]{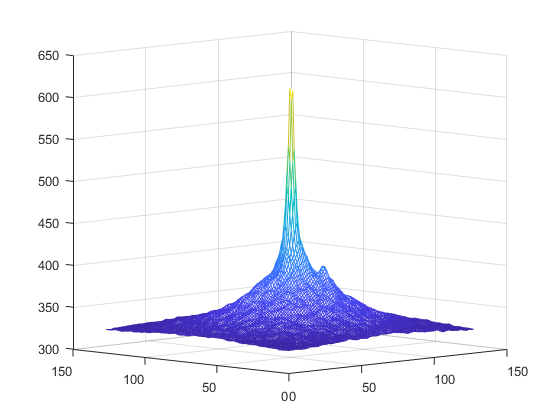}
  \end{minipage}%
  }

     \subfloat[LF.]{
  \label{fig2:subfig:d}
  \begin{minipage}[t]{0.5\linewidth}
  \centering
   \includegraphics[scale=0.8]{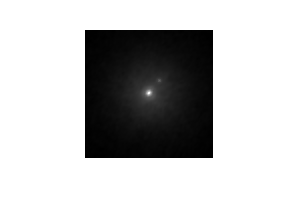}
  \end{minipage}%
  \begin{minipage}[t]{0.5\linewidth}
  \centering
   \includegraphics[scale=0.4]{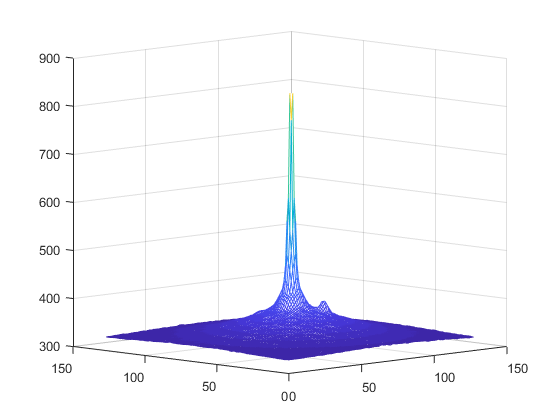}
  \end{minipage}%
  }
  
  \centering
  \caption{ Resultant images obtained by 4 different algorithm.}
  \label{Fig:fig2}
\end{figure}

Finally, Intuitive comparisons are made between the algorithm proposed in this paper and the space-domain algorithm and the frequency-domain algorithm by means of images and graphics respectively. Figure~\ref{fig2:subfig:a},~\ref{fig2:subfig:b},~\ref{fig2:subfig:c} and ~\ref{fig2:subfig:d} show resultant images obtained by the simple addition, the shift and addition, the LI and the LF, respectively. It can be seen intuitively from Figure~\ref{Fig:fig2} and Figure~\ref{Fig5} that with the continuous improvement of the processing method, the base of the binary becomes more and more flat, the companion becomes more and more clear, and the peak value becomes higher and higher, indicating that the algorithm proposed in this paper is better than the other space-domain algorithms and the frequency-domain algorithm.

\begin{figure}[ht]
   \centering
   \includegraphics[width=5in, angle=0]{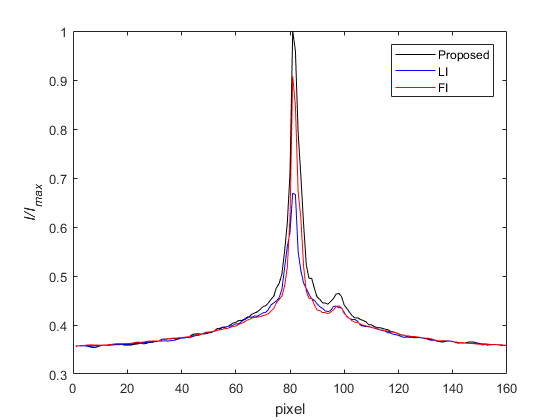}
   \caption{Profiles through the peaks of the primary and the companion obtained by 3 different algorithms. }
   \label{Fig8}
   \end{figure}

The three curves (i.e. the blue, the red and the black) in Figure~\ref{Fig8} show three cross-sections of the images obtained by the LI, the LF and the proposed algorithm respectively. From the perspective of the peak value of the primary star and the height of the companion star, the proposed algorithm is obviously better than the traditional LI and LF algorithm.
   
Therefore, no matter in the aspect of star image quality and of traditional image definition or of intuitive image and graphic display effect, the algorithm proposed in this paper is superior to the traditional LI and LF algorithms deigned and implemented by Mao and Hu, demonstrating as well that the iterative calculation with selection rate in this paper is effective. 

\subsection{Simulation Experiment of the Binary Star and Discussion}
\begin{figure}[ht]
\centering
  \subfloat{
  \begin{minipage}[t]{0.5\linewidth}
  \centering
   \includegraphics[scale=0.9]{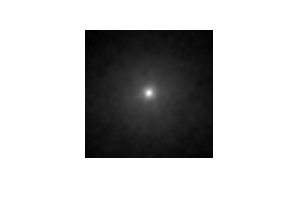}
  \end{minipage}%
  \begin{minipage}[t]{0.5\linewidth}
  \centering
   \includegraphics[scale=0.45]{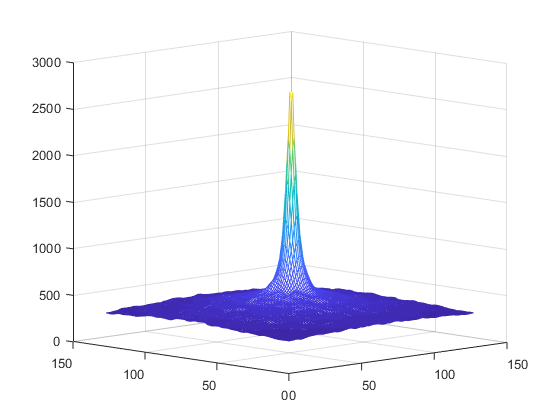}
  \end{minipage}%
  }
   \centering
  \caption{Resultant images of TOK 382.}
  \label{Fig9}
\end{figure}

The short-exposure images of the double star system TOK 382 are also processed in the same way as those of HDS70, and the lucky imaging result is shown in Figure~\ref{Fig9}. It can be seen from Figure~\ref{Fig9} that under the same observation conditions for HDS 70 (atmospheric coherent length $r_0$ of approximately 0.08$m$), unfortunately, it is unable to distinguish the faint companion of the binary by the hybrid algorithm, and the same applies to the LI and the LF algorithms, as is found through experiment.

\begin{figure}[ht]
\centering
  \subfloat[A random frame.]{
  \label{fig3:subfig:a}
  \begin{minipage}[t]{0.5\linewidth}
  \centering
   \includegraphics[scale=0.9]{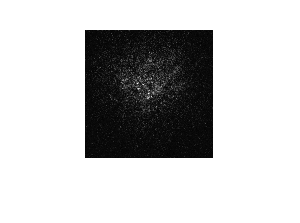}
  \end{minipage}%
  \begin{minipage}[t]{0.5\linewidth}
  \centering
   \includegraphics[scale=0.45]{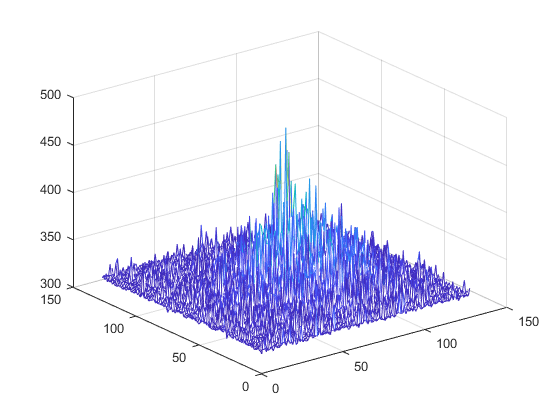}
  \end{minipage}%
  }
  
  \subfloat[Resultant image.]{
  \label{fig3:subfig:b}
  \begin{minipage}[t]{0.5\linewidth}
  \centering
   \includegraphics[scale=0.9]{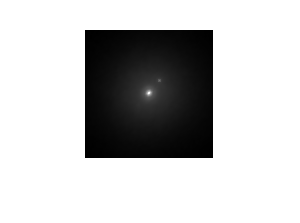}
  \end{minipage}%
  \begin{minipage}[t]{0.5\linewidth}
  \centering
   \includegraphics[scale=0.45]{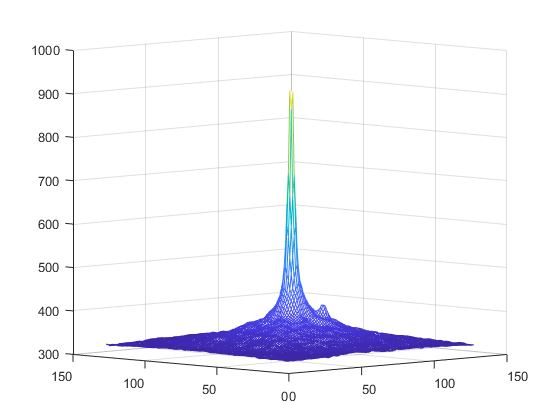}
  \end{minipage}%
  }
  \centering
  \caption{Short-exposure and resultant images of HDS 70 obtained through simulation.}
  \label{Fig:fig3}
\end{figure}

To explore the reason, we design a simulation experiment. First, by the method proposed by \cite{Yang+etal+2000} and on the basis of the numerical simulation program of imaging system developed by \cite{Yang+2016}, the relationship of digital numbers (DN) of images, the distribution of speckle areas and the major noise compositions (bias, dark current and its noise, photon shot noise) are determined through analysis of short-exposure observed images of HDS 70 and TOK 382, and a short-exposure image simulation program is designed and written to simulate the short-exposure observation by 2.4m-diameter telescope with EMCCD at Lijiang Observatory on the evening of Oct. 20, 2016. Multiple sets of short-exposure simulated images of HDS 70 at $r_0=0.08m$, each of 10,000 frames, are generated by this program and subjected to lucky imaging. The simulated short-exposure images and the resultant image, shown in Figure~\ref{Fig:fig3}, are very similar to the observed images of HDS 70 (shown in Figs~\ref{fig1:subfig:a} and~\ref{Fig5}, etc.), which indicates that the designed simulation program is feasible and effective. 

\begin{figure}[ht]
   \centering
   \includegraphics[width=5in, angle=0]{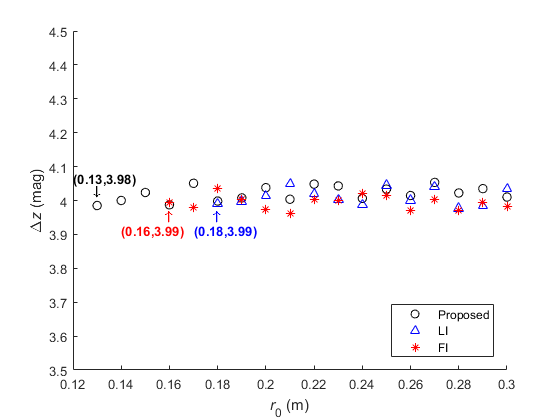}
   \caption{Detection capability at different $r_0$. }
   \label{Fig11}
   \end{figure}

Then, multiple sets of simulated images of the binary TOK 382 at different $r_0$, each of 10,000 frames, are generated by this simulation program and processed by the hybrid algorithm, the LI algorithm and the LF algorithm respectively, with the processing results shown in Figure~\ref{Fig11}. Figure~\ref{Fig11} illustrates the detection ability of the three algorithms from another perspective, in which abscissa $r_0$ ranges from 0.07m to 0.3m at an interval of 0.01m; ordinate $\Delta z$ is the magnitude difference between bright primary and faint companion estimated based on the resultant images obtained through these processing algorithms. The coordinate points given in Figure~\ref{Fig11} represent the minimum $r_0$ and the estimated magnitude difference when an algorithm can detect the companion star of TOK 382. For example, the black coordinate value (0.13, 3.98) indicates that the minimum $r_0$ is 0.13m and the magnitude difference is 3.98 when the proposed algorithm can detect the companion star of TOK 382. In other words, when $r_0$ is lower than 0.13m, this algorithm cannot separate the companion star from the bright spot of the primary star, and when $r_0$ is equal to or higher than 0.13m, the algorithm can detect the faint star of TOK 382. The other two given coordinate points also have the same meaning.

It can be seen from Figure~\ref{Fig11} that when $r_0$ is less than 0.13m, the faint star of the binary TOK 382 with magnitude difference of approximately 4mag and angular separation of approximately 0.4 arcsecond can be distinguished by none of these three algorithms. Under the atmospheric conditions of that very evening in Lijiang (seeing=1.4", $r_0$ of approx. 0.08m), it is unable to reconstruct a distinguishable image of the faint companion from the observation image of TOK 382 no matter which algorithm is used. Only when atmospheric conditions improve until coherent length reaches $r_0=0.13m$ can the faint star of the binary be distinguished by the hybrid algorithm, while for the LF algorithm and LI algorithm, the required coherent length is $r_0=0.16m$ and $r_0=0.18m$, respectively. This indicates from another perspective that the proposed hybrid algorithm is superior to the LI algorithm and the LF algorithm.

\begin{figure}[ht]
   \centering
   \includegraphics[width=5in, angle=0]{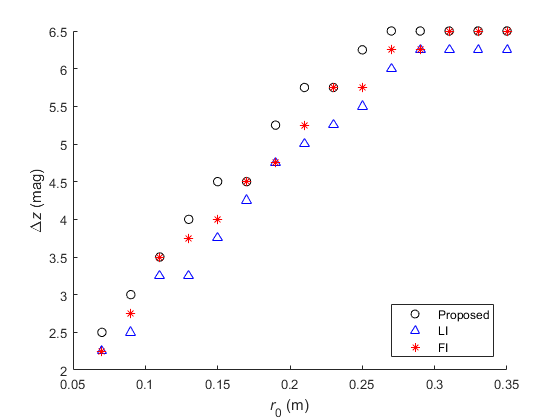}
   \caption{ Detection limits at different $r_0$. }
   \label{Fig12}
   \end{figure}

In order to test the detection limits of the three algorithms for binary stars under different atmospheric conditions, multiple sets of images of a simulated binary star system at different $r_0$ and different $\Delta z$, each of 10,000 frames, are generated randomly and processed by the hybrid algorithm, the LI algorithm and the LF algorithm respectively, with the processing results shown in Figure~\ref{Fig12}. In this simulation, the angular separation is a fixed value of 0.5 arcsecond; $r_0$ ranges from 0.07m to 0.3m at an interval of 0.02m and $\Delta z$ ranges from 0mag to 8mag at an interval of 0.25mag. The points with the same color in Figure~\ref{Fig12} indicate the maximum magnitude difference of a binary star when an algorithm can detect a faint companion star in the binary star under the different $r_0$. For example, when $r_0$ is 0.15m, the LI, the LF, and the proposed algorithm can detect a faint companion star in a binary star system with the largest magnitude difference of 3.75mag, 4mag, and 4.5mag, respectively, and other points have the same meaning.

As can be seen from Figure~\ref{Fig12}, in the case that the faint star of the binary can be distinguished by the three algorithms, the maximum magnitude differences increase with $r_0$ when $r_0$ is less than or equal to 0.27m while basically remaining stable when $r_0$ is greater than 0.27. At an angular separation of 0.5 arcsecond, the faint star of the binary with magnitude difference of at most 6.5mag can be distinguished by the hybrid algorithm and the LF algorithm and at most 6.25mag by the LI algorithm. This simulation result indicates a better sensitivity than the measured sensitivity of AstraLux(\citealt{Janson+etal+2012, Bergfors+etal+2013}) as there are more interference factors in actual observation. When $r_0$ is 0.08m and in the case that the faint star of the binary can be distinguished by the three algorithms, the maximum magnitude differences are all less than 4mag, further showing that under the atmospheric conditions of that very evening in Lijiang (seeing=1.4", $r_0$ of approx. 0.08m), it is unable to distinguish the faint star from the observation image of TOK 382 no matter which algorithm is used. Though we fail to distinguish the faint companion from the observation image of TOK 382 under current conditions, it can be seen from the distribution of the maximum magnitude differences in the case that the faint star of the binary can be distinguished by the three algorithms at different $r_0$ in Figure~\ref{Fig12} that the algorithm proposed in this paper is superior to the LI and LF algorithms. 

Based on the above simulation analysis, we believe that to obtain good data processing results, it is necessary to carry out lucky imaging observation at an observatory with good seeing or use a lucky imaging system equipped with a low-order AO system for observation, e.g. AOLI of the University of Cambridge(\citealt{Mackay+etal+2012}). Domestically, the seeing of astronomical observation sites in Muztagh Ata, Xinjiang and Daocheng, Sichuan of western China reaches 0.4 and 0.5 acrsecond, with median of 0.7 and 0.9 arcsecond respectively(\citealt{Cao+etal+2020, Song+etal+2020, Xu+etal+2020}). After establishment of these two observatories, we can probably expect according to the simulation result in Figure~\ref{Fig12} that the faint star of TOK 382 can be distinguished through lucky imaging observation with 2m-class telescope and EMCCD camera under normal atmospheric conditions in Muztagh Ata or Daocheng; and if atmospheric conditions are better, the faint star of the binary with magnitude difference of 6 mag and angular separation of 0.5 arcsecond can be distinguished by the lucky imaging algorithm proposed in this paper, achieving the level of AstraLux. 

To sum up, the hybrid algorithm proposed in this paper is superior to the conventional LI and the LF algorithms. Actually, the proposed algorithm incorporates these two lucky imaging algorithms as proper subsets. It can be clearly seen from the final high-resolution reconstructed image that the imaging effect of the proposed algorithm is obviously superior to that of the conventional LI and the LF algorithms designed and implemented by Mao and by Hu respectively. This algorithm, however, needs improvement in selection and optimization of parameters $r_s$, $d_s$, $r_f$ and $d_f$, of which too small values may lead to a very long calculation time while too large values may lead to decreased imaging effect. One possible solution is to use GPU to process some tasks that require a large amount of calculation, e.g. FFT and IFFT.

\section{Conclusions}
\label{sect:conclusion}
Based on a careful analysis of the lucky imaging methods studied by Mackay and by Garrel et al and the lucky imaging algorithms designed and implemented by Mao and by Hu et al, this paper proposes a novel hybrid algorithm of LI and LF where, with space-domain and frequency-domain selection rates as a link, the two classic lucky imaging algorithms are combined as proper subsets, space-domain and frequency-domain selection rates are made an increasing sequence instead of fixed values, data are selected and superposed in many levels and aspects according to certain rules (e.g. the instantaneous Strehl ratio for LI algorithm and the amplitude for LF algorithm), and through fusion of various results of selection and superposition, the final high-resolution reconstructed image is obtained, of which FWHM is approximately 0.127", superior to that obtained by the two classic lucky imaging algorithms. Besides, this paper also proposes a new selection and storage scheme for lucky imaging algorithm, which can significantly save computer memory and enable lucky imaging algorithm to be implemented in a common desktop or laptop with small memory and to process astronomical images with more frames and larger size. Furthermore, it is found through simulation analysis that with better atmospheric seeing, the imaging resolution of both the novel and the traditional lucky imaging algorithms can be improved, but reaches the limit at the inflection point of $r_0=0.27m$; under the same atmospheric conditions, the imaging resolution of the binary achieved by the novel hybrid algorithm is superior to both that by the classic LI algorithm and that by the LF algorithm. 

\begin{acknowledgements}
Thanks to Prof. K. Ji and Prof. Z. Jin from Yunnan Observatory, Chinese Academy of Sciences for their valuable comments and suggestions for this study. This work is supported by the National Natural Science Foundation of China under Grant 11673009.
\end{acknowledgements}

\label{lastpage}

\end{document}